\documentclass[twocolumn,showpacs,amsmath,amssymb]{revtex4}

\usepackage{graphicx}
\usepackage{dcolumn}
\usepackage{bm}

\begin{document}


\title{Structure of Stochastic Dynamics near Fixed Points}

\author{Chulan Kwon}
\email{ckwon@mju.ac.kr}
 \affiliation{Department of Physics, Myongji University,
   Namdong San 38-2, Yongin, Kyonggi-Do, 449-728, Republic of
   Korea}
\author{Ping Ao}%
 \email{aoping@u.washington.edu}
\affiliation{ Department of Mechanical Engineering, University of
Washington, Seattle, WA 98195, USA}%
\author{David J. Thouless}
 \email{thouless@u.washington.edu}
\affiliation{Department of Physics, University of Washington,  Seattle, WA 98195, USA}%

\date{\today}

\begin{abstract}
We analyze the structure of stochastic dynamics near either a
stable or unstable fixed point, where force can be approximated by
linearization. We find that a cost function that determines a
Boltzmann-like stationary distribution can always be defined near
it. Such a stationary distribution does not need to satisfy the
usual detailed balance condition, but might have instead a
divergence-free probability current.  In the linear case the force
can be split into two parts, one of which gives detailed balance
with the diffusive motion, while the other induces cyclic motion
on surfaces of constant cost function. Using the Jordan
transformation for the force matrix, we find an explicit
construction of the cost function. We discuss singularities of the
transformation and their consequences for the stationary
distribution.  This Boltzmann-like distribution may be not unique,
and nonlinear effects and boundary conditions may change the
distribution and induce additional currents even in the
neighborhood of a fixed point.
\end{abstract}

\maketitle

\section{ Introduction }

  In equilibrium statistical mechanics an important role is played by the
principle of detailed balance and by the related
fluctuation--dissipation theorem.  Einstein used the principle
that the excess energy that is put into each mode of an
equilibrium system in the course of thermal fluctuations is also
removed from the same mode by dissipative forces. This is implicit
in his work on Brownian movement \cite{einstein05}, and explicit
in later works on the photoelectric effect \cite{einstein12}, and
on the relation between spontaneous and induced emission of
electromagnetic radiation \cite{einstein17}.  This principle was
formulated as the principle of detailed balance by Bridgman
\cite{bridgman28}, and used to explain Johnson noise in electrical
circuits by Nyquist \cite{nyquist28}.  This is related to the fact
that the same processes that drive fluctuations in the
neighborhood of a typical equilibrium configuration also drive the
configuration back towards a typical equilibrium or steady state
configuration when it is displaced from equilibrium by an amount
which is small, but large compared with the fluctuations in
thermal equilibrium.  In this situation the equilibrium
distribution in phase space is just the Boltzmann distribution,
proportional to $e^{-\beta E}$, where $\beta$ is inversely
proportional to temperature, and $E$ is the energy of the point in
phase space.  Configurations that differ significantly from those
that contribute to the minimum of the free energy are driven back
to the neighborhood of this minimum by dissipative effects such as
thermal or electrical conduction, or viscosity, and the magnitude
of these effects is related to the equilibrium fluctuations of
related variables.

  In many situations there is no thermodynamic equilibrium, but external
steady and fluctuating forces drive the system into a steady or
very slowly varying state for which the principle of detailed
balance does not hold.  A light bulb powered by an external
battery, or a chemical reaction in which the reactants are
introduced at a steady rate and the products of the reaction
removed at a steady rate, would both be examples of such a
situation.  Even in a situation which is almost in equilibrium,
such as a system which is started in equilibrium at a local
minimum of the free energy, but which can can go over a saddle
point to a deeper minimum, the behavior near the saddle point does
not satisfy the principle of detailed balance, since there is a
current over the saddle point.

  For such systems without detailed balance there is no general method of
obtaining the equilibrium distribution from a knowledge of the
steady and stochastic forces, such as the Boltzmann distribution
provides for a system with detailed balance.  In recent work one
of us \cite{ao03} has developed a method valid near a stable fixed
point which, even when detailed balance does not hold, obtains a
cost function analogous to the energy for the Boltzmann
distribution.  If this method can be extended away from the linear
region in the neighborhood of a fixed point it may provide a new
method for dealing with problems of this sort \cite{zhu03}.

 Great efforts have been spent on finding such a cost function ever
since the work of Onsager \cite{onsager}. Results up to 1990 have
been summarized, for example, by van Kampen \cite{vankampen}.
%
In general such efforts have been regarded as not very successful
\cite{cross}.

  In spite of the difficulty, there have been continuous efforts on
the construction of cost function and related topics. Elegant
results have been obtained in several directions. Tanase-Nicola
and Kurchan \cite{kurchan} have considered explicitly the saddle
points of gradient systems. They started from the existence of
potential or cost function to avoid the most difficult problem of
the irreversibility. The gain is that they can now obtain a
powerful computational method to count the saddle points and to
compute the escape rate. They also provide an extensive list of
related literature.

 The study on the mismatch of the fixed points of the drift force
and the extremals of the steady state distribution has been
reviewed by Lindner {\it et al.} \cite{lindner}. Rich phenomena
have been observed, but the mismatch has been treated as
``experimental'' result. There is no mathematical/theoretical
explanation on why it should happen.

 In another survey the useful and constructive role played by the
noise has been demonstrated by examples \cite{wio}. It is argued
that the noise is essential to establish the functions of
dynamical systems. Again, the mismatch problem is encountered and
the constructed potential function is often regarded as
approximation.

 From a different perspective, there has been an effort to provide a solid
foundation for non-equilibrium processes based on the chaotic
hypothesis \cite{gbg}.  The chaotic hypothesis presumes that the
system is sufficiently chaotic that variation of parameters of the
system leads to a unique parameter-dependent steady state, even
though the Gibbs entropy change is not path
independent\cite{ruelle03}.  Under this hypothesis an interesting
and important fluctuation theorem has been obtained, which further
suggests the existence of the Boltzmann-like steady state
distribution function. Hence, a cost function very likely exists
under this situation. A difficulty with this approach is that
extremely few practical physical systems have been shown to
satisfy the chaotic hypothesis.

 Because the metastability is such an important phenomenon and
because of the difficulty encountered in the construction of cost
function, efforts have been made to go around the cost function
problem when computing the life time of a metastable state.  The
effort results in the so-called Machlup-Onsager functional method,
summarized by Freidlin and Wentzell \cite{freidlin}. This approach
has been actively pursued recently \cite{maier,beri}.

 In this paper we give a careful discussion of the basis of our method
of constructing a cost function in the linear region close to any
fixed point, whether stable or unstable.  We show that this
approach gives an unambiguous prescription under wide conditions.
The only case we have found which does not give an unambiguous
expression for the cost function has a subspace within which the
noise does not act, and out of which the force does not carry the
state.

 In section \ref{section2} we give a general discussion of a linear
system with noise.  In section \ref{section3} we show how a cost
function can be constructed by a decomposition of the force matrix
into two factors, one of which is a symmetric cost function
matrix, and that this gives a probability density of exponential
form, with the exponent proportional to the cost function.  The
general proof of this result is obtained in the Appendix, where we
exploit the Jordan transformation for matrices with an incomplete
set of eigenvectors.  In section \ref{section4} we discuss the
singularities of the transformation, and identify two types of
singularities, one of which corresponds to a flat subspace of the
cost function, while the other corresponds to the possibility that
the dynamics separates the system into two or more disjoint
subspaces.  In section \ref{nonlinear} we discuss solutions other
than the Boltzmann-like solution of the equation for a stationary
distribution, and argue that such solutions may be significant in
any attempt to extend this solution to the nonlinear regime.   In
section \ref{discussion} we discuss the significance of this
decomposition of the force matrix, and its relation to the
principle of detailed balance.

\section{ Stating the Problem }
\label{section2}

 Many processes in natural sciences can be modeled quantitatively. One
particularly important class of such modeling is that described by
first order differential equations \cite{kaplan}, supplemented by
stochastic terms \cite{vankampen}.  We start with the nonlinear
dynamic equation
\begin{equation}
   \dot{x}= f(x) + \zeta(x,t) \;, \label{langevin}
\end{equation}
 which gives the stochastic evolution of the state represented by
the real $d$-dimensional vector, $ x^{\tau} = (x_1, ... , x_d )$.
Here the superscript $\tau$ denotes the transposed vector.  The
force vector is the $d$-component $f(x)$, and this gives the
deterministic time evolution of the system.  For simplicity we
take the noise term $\zeta(x,t)$ to be Gaussian white noise, with
zero mean, $\langle \zeta(t)\rangle = 0$, and variance $ \langle
\zeta(x,t) \zeta^{\tau}(x,t') \rangle = 2 D(x) \delta (t-t')$.
The  angular brackets denote the average over noise distribution,
and $\delta(t)$ is the Dirac delta function.  In this work we
assume $D(x)$ to be independent of $x$.  The probability
distribution function $\rho(x,t)$ then satisfies the Fokker-Planck
equation
\begin{equation} \label{eq:fokker}
 {{\partial \rho}(x,t) \over {\partial t}}
 = {{\partial }\over{\partial x_{i}}} \left[-f_{i}(x)
   +D_{ij}{\partial\over  \partial x_{j}}\right]\rho(x,t)~.
\end{equation}

  In the neighborhood of a fixed point, which we take to be at the origin,
the force can be replaced by its linear approximation
\begin{equation}\label{eq:linearf}
f_i(x)=F_{ij}x_j\;. \end{equation}
 It was noticed by Ao \cite{ao03} that, in the linear region near a stable
fixed point, Eqs.\ (\ref{langevin}) and (\ref{eq:linearf}) can be
decomposed in the form
\begin{equation}
  ( S + A ) \dot{x} = - U x + \xi(t) \; ,\label{gauge}
\end{equation}
 where the symmetric matrix $S$ is semi-positive definite and the matrix
$A$ is antisymmetric.  The noise function $\xi$ has variance given
by $ \langle \xi(x,t) \xi^{\tau}(x,t') \rangle = 2 S \delta
(t-t')$.  In this paper we adopt an equivalent, but simpler,
approach of factorizing the force matrix as
\begin{equation}\label{eq:factors}
F= -(D+Q)U= -(S+A)^{-1}U \;,\end{equation}
 where $D$ is the symmetric diffusion matrix, $Q$ is an antisymmetric
matrix which can be determined, and $U$ is the symmetric cost
function matrix, which was called a potential matrix in ref.\
\ref{ref:ao03}. This breaks the force $Fx$ into two components,
$F^{(d)}x=-DUx$, which generates a motion towards the origin if
$U$ is positive definite, and $F^{(c)}x=-QUx$, which gives a
motion on the manifold of constant ${\cal U}(x)=(1/2)x^\tau Ux$.
The quadratic form ${\cal U}(x)$ is the cost function.

 It is immediately obvious that if the vector $f$ is replaced by
$F^{(d)}x=-DUx$ in Eq.\ (\ref{eq:fokker}), a Boltzmann-like
stationary distribution of the form:
\begin{equation}
   \rho( x ) \propto \exp\{ - {\cal U}(x) \} \label{stationary}
\end{equation}
satisfies the equation, since the current density
\begin{equation}\label{eq:current1}
j_i(x)=\left[f_i(x)-D_{ij}\partial/\partial x_j\right]\rho
\end{equation}
vanishes.  Since $F^{(c)}x$ generates a current density
\begin{equation}\label{eq:current2}
j^{(c)}(x)=-QUx\rho(x)\;,
\end{equation}
which is divergence free, and conserves $\cal U$ and $\rho$, the
combination $f(x)=[F^{(d)}+F^{(c)}]x$  also conserves the
distribution given by Eq.\ (\ref{stationary}), so that this is a
stationary solution of Eq.\ (\ref{eq:fokker}).

  This decomposition of the force matrix allows an explicit
time-independent solution of the Fokker-Planck equation to be
written down.  In ref.\ \ref{ref:ao03} the solution of the
equation for $Q$ was obtained by a power series expansion, without
any discussion of the convergence of this series.  In the next
section we show that there is a unique solution for the equation
for $Q$ under rather wide conditions. It is not even required that
the fixed point of $F$ be stable, although if it is not the
stationary solution given by Eq.\ (\ref{stationary}) is unbounded,
and could only, at best, give a useful solution in a neighborhood
of the fixed points with boundary conditions that do not perturb
this solution too strongly.

 It is worth noting that, under the coordinate transformation $x\to
y=M^{-1}x$, the mapping $F$ that relates $\dot x$ to $x$
transforms as
\begin{equation}\label{eq:transformf}
F\to M^{-1}FM \;,\end{equation} while the symmetric matrices $U$
and $D$, which represent quadratic forms, transform as
\begin{equation}\label{eq:transformud}
U\to M^\tau UM \;,\ \ D\to M^{-1} D\left(M^\tau\right)^{-1}
\;,\end{equation} and $Q$ transforms in the same way as $D$. These
transformations preserve the separation of $D+Q$ into symmetric
and antisymmetric parts.

 Starting from the work of Onsager \cite{onsager}, there has been an
extensive literature on dynamical behavior near a stable fixed
point \cite{sampling}.  The new construction clearly works for
this important situation, and has indeed offered a new angle.
However, it is not sufficient to generalize the decomposition of
the force in Eq.\ (\ref{eq:factors}) to the nonlinear regime by an
equation of the form $f=-(D+Q){\rm grad}\,\cal U$, since the
Boltzmann form, Eq. (\ref{stationary}) may not satisfy the
Fokker-Planck equation ({\ref{eq:fokker}) if the antisymmetric
matrix $Q$ is space dependent. The generalization to nonlinear
systems needs further study. In the next three sections we
establish the decomposition firmly in the linear regime, and
investigate its limitations and implications.

\section{Decomposition of the Force Matrix }
\label{section3}

  In this section we develop a general method for making the decomposition
of the force matrix given in Eq.\ (\ref{eq:factors}).  Since this
equation, together with the symmetry of $U,D$, and antisymmetry of
$Q$, leads to
\begin{equation}
  U = - (D+Q)^{-1} F= -F^\tau(D-Q)^{-1}\;, \label{rel1}
\end{equation}
the equation to determine $Q$ is
\begin{equation}
   FQ+QF^{\tau} = FD-DF^{\tau}\;. \label{rel2}
\end{equation}
This is a system of $d(d-1)/2$ linear equations to determine the
same number of independent components of $Q$, so it has a unique
solution unless the set of equations is singular.  Inversion of
the matrix $D+Q$ then gives the matrix $S+A$ of Eq.\
(\ref{gauge}).

  Our method of solution is best illustrated by considering the case that
$F$ is real symmetric or has distinct eigenvalues, so that it can
be diagonalized in terms of its left and right eigenvectors.
Equation (\ref{rel2}) then takes the form
\begin{equation}\label{eq:fdiag}
(\lambda_\alpha+\lambda_\beta)\tilde Q_{\alpha\beta}=
(\lambda_\alpha-\lambda_\beta)\tilde D_{\alpha\beta}
\;,\end{equation} where the $\lambda_\alpha$ are the eigenvalues
of $F$, and the tilde denotes this representation in terms of
eigenvectors.  This gives an immediate solution for $Q$ provided
that no pair of the eigenvalues of $F$ adds up to zero.  The
eigenvalues can only add to zero when the fixed point is unstable,
and this is discussed in section \ref{section4}.

 For completeness we must consider the general case with degenerate
eigenvalues for asymmetric $F$, in which case there may not be a
complete set of eigenvectors. This case is dealt with in the
Appendix, using the Jordan representation of a nonsymmetric
matrix.

\section{Singularities of the Transformation}
\label{section4}

 There are two places in our argument where the transformation from
the force matrix $F$ to the symmetric cost function matrix $U$
might be singular. Equation (\ref{rel2}) for the antisymmetric
matrix $Q$ can be solved, and we have an explicit solution in
section \ref{section3} unless the determinant of the coefficients
in $d(d-1)/2$ inhomogeneous equations is zero. The second
possibility is that the matrix $D+Q$ whose inverse appears in Eq.\
(\ref{rel1}) might have zero determinant.

 In section \ref{section3} we showed that the conditions for the
equation for $Q$ to be singular are that two of the eigenvalues of
$F$ sum to zero, or, as can be seen from Eq.\
(\ref{eq:jordandiag}) in the Appendix, when the null space of
$F^2$ has two or more dimensions. Neither of these cases arise for
a stable fixed point.  There are two distinct cases of
$\lambda_\alpha+\lambda_\beta=0$, according to whether the two
eigenvalues are real eigenvalues of opposite sign, or whether they
form a complex conjugate pair. We study the behavior of the
eigenvalues and eigenvectors of $U$ in these two cases, assuming
that the two eigenvalues of $F$ are nondegenerate, and that $F$
has no zero eigenvalue.

 Instead of studying the eigenvectors of $U$ directly, we study the
eigenvalues and eigenvectors of
\begin{equation}\label{eq:uinv}
U^{-1}=-F^{-1}(D+Q)=-R\Lambda^{-1}(\tilde D+\tilde Q)R^\tau \;,
\end{equation}
 which has the same eigenvectors but reciprocal eigenvalues. Here $R$ is
the matrix whose columns are the right eigenvectors of $F$, and
$\Lambda$ is the diagonal matrix with the eigenvalues of $F$ as
its diagonal elements.  The generalization of these definitions of
$R$ and $\Lambda$ to the case where the eigenvalues of $F$ are not
complete is given in the Appendix in Eqs.\ (\ref{eq:lambda}),
(\ref{eq:tilde}) and (\ref{eq:utilde}). For a pair of eigenvalues
with $\lambda_\alpha+\lambda_\beta\approx 0$, with no other sums
of two eigenvalues small and no other small individual
eigenvalues, the only large terms in the matrix $\tilde
U^{-1}=-\Lambda^{-1}(\tilde D+\tilde Q)$ are, according to Eq.\
(\ref{eq:fdiag}),
\begin{equation} (\tilde U^{-1})_{\alpha\beta} =(\tilde
 U^{-1})_{\beta\alpha} =-{2\tilde D_{\alpha\beta} \over
\lambda_\alpha+\lambda_\beta} \;.
\end{equation}
 When all other matrix elements are neglected, this approximation,
combined with Eq.\ (\ref{eq:uinv}), gives
\begin{equation}
 \sum_j\left(U^{-1}\right)_{ij}L_{\gamma j}\approx  -{2\tilde
D_{\alpha\beta}
 \over \lambda_\alpha+\lambda_\beta} \left(\delta_{\gamma\alpha}
 R_{i\beta} +\delta_{\gamma\beta}
 R_{i\alpha}\right) \;.
\end{equation}
This equation, combined with the relation $RL=I$, shows that the
approximate eigenvector corresponding to a large eigenvalue
$w^{-1}$ can be written as
\begin{equation}
 a_\alpha R_{i\alpha}+a_\beta R_{i\beta}
 =\sum_j(a_\alpha R_{j\alpha}+a_\beta R_{j\beta})\sum_\gamma
 R_{j\gamma}L_{\gamma i} \;,
\end{equation}
provided the amplitudes and eigenvalues satisfy the equation
\begin{equation}
 \begin{array}{cc}
  w^{-1} a_\alpha &= -{2\tilde D_{\alpha\beta} \over
   \lambda_\alpha+\lambda_\beta}\left(a_\alpha\sum_j
   R_{j\alpha}R_{j\beta} +a_\beta\sum_j R_{j\beta}^2 \right) \;, \\
  w^{-1} a_\beta &= -{2\tilde D_{\alpha\beta} \over
   \lambda_\alpha+\lambda_\beta}\left(a_\alpha\sum_j R_{j\alpha}^2
   +a_\beta\sum_j R_{j\alpha}R_{j\beta} \right)\;.
  \end{array}
\end{equation}
This gives the two small real eigenvalues of $U$ as
\begin{equation}
  w \approx -{\lambda_\alpha+\lambda_\beta \over 2\tilde
  D_{\alpha\beta}}\left( \sum_j R_{j\alpha} R_{j\beta}\pm
  \sqrt{\sum_i R_{i\alpha}^2 \sum_j R_{j\beta}^2} \right)^{-1} \;,
\label{eigenvalue}
\end{equation}
and the corresponding eigenvectors as
\begin{equation} {a_\alpha\over a_\beta}\approx \pm \sqrt{\sum_j
R_{j\beta}^2\over \sum_j R_{j\alpha}^2 } \;.
\end{equation}

 For the case of a pair of real eigenvalues of opposite signs we can
see that, as the sign of $\lambda_\alpha+\lambda_\beta$ is changed
by a change in the parameters of $F,D,$ the stable and unstable
manifolds of $U$ change places with one another.  The stable and
unstable manifolds of $U$ bisect the stable and unstable manifolds
of $F$ in the original representation, as is obvious if one
normalizes the real eigenvectors of $\tilde U$ by $ {\hat
R}_{i\alpha}=R_{i\alpha}/\sqrt{\sum_j R_{j\alpha}^{2}}$ and ${\hat
R}_{i\beta}$ similarly, giving eigenvectors, ${\hat
R}_{i\alpha}\pm {\hat R}_{i\beta}$. In the limit
$\lambda_\alpha=\lambda_\beta$, $U$ is flat in this
two-dimensional subspace.

 For a complex conjugate pair of eigenvalues with
$\lambda_\alpha+\lambda_\alpha^*\approx 0$ the behavior is a
little different. Using the property $R_{j\alpha}=R_{j\beta}^*$,
we find the term inside the bracket in Eq.~(\ref{eigenvalue}) is
always positive. The two eigenvalues of $U$ then have the same
sign, so $U$ is either stable or unstable, depending on the sign
of $\lambda_{\alpha}+\lambda_{\beta}$, in this two dimensional
subspace. As the real part of $\lambda_\alpha$ changes sign, a
two-dimensional stable manifold becomes unstable, or vice versa.

 It can be seen from Eq.\ (\ref{rel3}) in the Appendix that where one of
the eigenvalues satisfying $\lambda_\alpha+\lambda_\beta=0$
corresponds to a higher dimensional subspace there may be higher
order zeros of the eigenvalues of $U$.

 If the matrix $D$ is positive definite there is no possibility that
$D+Q$ could be singular. If $ u$ is a vector in the null space of
$D+Q$, we have
\begin{equation}\label{eq:expect}
0= u^\tau (D+Q)u =u^\tau D u\;,
\end{equation}
 since the antisymmetry of $Q$ makes its expectation value vanish.
Therefore $D\pm Q$ cannot be singular when $D$ is positive
definite.

 However, we do not usually want to specify that the noise acts on all
coordinates.  Typically, when two of the coordinates are the
position and momentum of a particle, people will take the noise to
change the momentum but not the position of the particle.
However, Eq.\ (\ref{eq:expect}) shows that for non-negative
definite $D$, vectors in the null space of $D+Q$ are in the
intersection of the null spaces of $D$ and $Q$. Equation
(\ref{rel2}) then shows that, for such a vector $u$ in the null
space of $D$ and $Q$,
\begin{equation}
  0=F(D-Q)u=(D+Q)F^\tau u\;,
\end{equation}
 and so $u$ is only in this null space if $F^\tau u$, or any power of
$F^\tau$ acting on $u$, is still in the null space.

 This condition is in agreement with what one should expect.  The noise
does not have to act directly on all coordinates, but, if there is
a subspace in which there is no noise, and which is left invariant
by the motion, there can be no equilibration within that subspace
except collapse towards a stable fixed point.

\section{Other Stationary Solutions}\label{nonlinear}

Although the Boltzmann-like form given in Eq.\ (\ref{stationary})
gives a stationary solution of the Fokker-Planck equation, it is
only the unique solution under certain rather restrictive boundary
conditions. One can see clearly why this might be an issue by
considering the one-dimensional form of the equation near a stable
fixed point, which can be written as
\begin{equation}\label{eq:1dstable} {d^2\rho\over dx^2}+{d\over
dx}(x\rho)=0\;. \end{equation} In addition to the Boltzmann-like
solution $\rho^{(0)}=(2\pi)^{-1/2}\exp(-x^2/2)$, this has a
current-carrying solution of the form
\begin{equation}\label{eq:1dstablej} \rho^{(1)}  (x)\propto
e^{-x^2/2}\int_0^x e^{x'^2/2}dx'\;. \end{equation} This expression
is proportional to $1/x$ for large values of $x$, so, if the
linear approximation to the equation is valid up to fairly large
values of $x$, the coefficient of such a term must be
exponentially small to prevent the probability density given by
$\rho^{(0)}+\rho^{(1)}$ from being negative.

  In $d$-dimensional systems there are similar solutions falling off
like $1/|x|^d$ for large $|x|$.  For such solutions of Eq.\
(\ref{eq:fokker}) the current at large distances from the origin
is primarily driven by the linear force $Fx$, and the diffusive
motion is a small correction, so, while the density falls off like
$|x|^{-d}$, the conserved current falls off like $|x|^{-d+1}$.
Again, current conservation shows that this contribution to the
density must be positive and negative in different parts of space,
so that its coefficient at the origin must be exponentially small,
with an exponent that depends on the size of the region in which
the linear approximation is valid.

  Near a maximum of the cost function, where
$\rho^{(0)}\propto\exp(x^2/2)$, the one-dimensional
current-carrying solution has the form
\begin{equation}\label{eq:1dunstablej} \rho^{(1)}(x)\propto
e^{x^2/2}\int_0^x e^{-x'^2/2}dx'\;. \end{equation} This grows at
large distances in the same way as $\rho^{(0)}$, with a change of
sign at the origin.  In $d$ dimensions a saddle point can sustain
a relatively large current across it, because there are
current-carrying states for which $\rho$ is of the same order of
magnitude as $\rho^{(0)}$.

  To get such a current across a stationary point in a linear system, it
is necessary to impose some external current sources and sinks.
However, if we want to describe a nonlinear force field in terms
of its approximately linear behavior in the neighborhood of its
fixed points, adjacent neighborhoods can generate external current
sources and sinks for one another, so we should not be surprised
to find such currents if we linearize in a local region.  These
``external'' currents will not only produce flows at the
boundaries, but will shift the flow lines away from the surfaces
of constant cost function $U$ shown in Eq.\ (\ref{eq:current2}).

  In the neighborhood of the minimum of the cost function, such
current-carrying solutions resembling the one-dimensional Eq.\
(\ref{eq:1dstablej}) shift the maximum of the density away from
fixed point, since the gradient of $\rho^{(1)}$ is nonzero.
Since, as we remarked in connection with this equation, the
amplitude of such a term must fall off exponentially with the size
of the region of linearization, in order to prevent negative
densities, it should not be possible to obtain such a term by a
conventional perturbation theory in the neighborhood of the stable
fixed point.  Our numerical exploration of nonlinear systems of
this sort suggests that these current-carrying solutions are
significant, since the maximum of the density is displaced from
the zero of the force.  One possibility is to introduce such
current-carrying states, in addition to a $\rho^{(0)}$ determined
by the cost function, in order to make this method applicable to
nonlinear systems.

\section{Discussion}\label{discussion}

  The main result of this work is to show that, for a system with a
deterministic motion controlled by a linear force, and a diffusive
motion driven by constant white noise, the force matrix $F$ can be
decomposed into two parts, $-DU$ and $-QU$.  Provided the fixed
point of $F$ is stable, the first of these components leads to a
steady state distribution of the Boltzmann form, $\exp(-{\cal
U}(x))$, with no probability current, the usual form of an
equilibrium distribution when detailed balance holds.  The second
part gives a flow on the surface of constant $\cal U$.

  The cost function matrix $U$ can be diagonalized by an orthogonal
transformation, and if it is positive definite, it can be
transformed to the identity matrix by choosing a new scale for the
variables.  In this representation the dissipative part of the
force matrix is $F^{(d)}=-D$, which is suggestive of Einstein's
relation between diffusion and dissipation \cite{einstein05}, or
of the fluctuation--dissipation theorem \cite{callen}.

  When the cyclic motion induced by $F^{(c)}$ is included the relation
between the eigenvalues of $F$ and $D$ becomes more complicated.
If the motion in a two-dimensional subspace is dominated by a fast
cyclic motion within the subspace, there will be a complex
conjugate pair of eigenvalues of $F$, so that they have a common
relaxation rate given by the real part of the eigenvalues.  These
possibilities require much more detailed work than we have yet
given them.

  This is actually not so different from the situation usually encountered
in statistical mechanics, at least if a classical rather than a
quantum description is used.  For a damped harmonic oscillator
there is a cyclic motion in phase space, as well as the thermal
noise and viscous damping acting on the momentum coordinate, while
for black-body radiation in a cavity there is cyclic motion
between electric and magnetic fields, in addition to the resistive
damping and noise from the walls of the cavity.

  What is remarkable is not that the steady state density can be written
as the exponential of a cost function, since if there is a steady
state we could always define the cost function as minus the
logarithm of the steady state density.  We find it remarkable that
for a linear stochastic system of this sort it is generally true
that the force can be decomposed into two parts, one of which
gives detailed balance in its strictest sense, while the other
gives a cyclic motion on the surfaces of constant cost function.

  The question of whether this technique can be extended to
nonlinear systems is an important one, but requires careful
investigation.  We have done preliminary work based on
perturbative inclusion of nonlinear terms, and made comparison
with numerical calculations.  It is clear that Eqs.\ (\ref{gauge})
and (\ref{eq:factors}) need some modification for nonlinear
effects, and also that if we try to solve the problem by matching
limited regions in which linearity holds approximately, the
matching may introduce solutions of the linearized Fokker-Planck
equation other than the Boltzmann-like $\rho^{(0)}$.  Different
regions will have to serve as sources and sinks for their adjacent
regions.

\begin{acknowledgments}
Discussions with M. Cross, S.W. Rhee, L. Yin, and X.-M. Zhu are
highly appreciated.  We are particularly grateful to E. Siggia,
whose criticisms of this paper have led us to make some important
modifications.  This work was supported in part by Institute for
Systems Biology (P.A.), by the National Institutes of Health
through Grant number HG002894, and by the National Science
Foundation through Grant number DMR-0201948.
\end{acknowledgments}

\appendix

\section{Appendixes}

For cases in which the force matrix $F$ has degenerate eigenvalues
and is not symmetric, so that it may not have a complete set of
eigenvectors, we use the Jordan transformation \cite{hirsch} of a
general real matrix. The Jordan transformation uses a complete set
of independent column vectors $v^\alpha$ with the property
\begin{equation}
   F v^\alpha=\lambda_\alpha v^\alpha+\mu_{\alpha-1}v^{\alpha-1} ~,
\end{equation}
 where $\mu_\alpha$ is zero if $\lambda_\alpha \ne \lambda_{\alpha+1}$
and is either unity or zero for $\lambda_\alpha =
\lambda_{\alpha+1}$. The set of row vectors $u^\alpha$ with the
orthonormality property $u^\alpha v^\beta=\delta_{\alpha\beta}$
then satisfies the equation
\begin{equation}
   u^\alpha F=\lambda_\alpha u^\alpha+\mu_{\alpha} u^{\alpha+1} ~.
\end{equation}
 Let us define matrices $R$ and $L$ as $R_{i\alpha}=v^{\alpha}_{i}$ and
 $L_{\alpha i}=u^{\alpha}_{i}$. Since  the vectors $u^\alpha$ are
orthonormal to the set $v^\alpha$, we have $LR=I$, so that these
matrices are inverse to one another. $R_{i\alpha}$, $L_{\alpha i}$
are real for real $\lambda_{\alpha}$. For complex
$\lambda_{\alpha}$, its complex conjugate is also an eigenvalue,
$\lambda_{\beta}=\lambda_{\alpha}^*$, since $F$ is a real matrix.
In this case $R_{\alpha i}=R_{\beta i}^*$,
$L_{i\alpha}=L_{i\beta}^*$. The Jordan transformation is then
given by
\begin{equation}\label{eq:lambda}
  LFR = \Lambda~.
\end{equation}
 The matrix $\Lambda$ is now block diagonal, where each nonzero block
has identical diagonal elements, which are degenerate eigenvalues,
and unity in each place immediately above the diagonal.

With this result, Eq.~(\ref{rel2}) can be rewritten as
\begin{equation}\label{rel4}
  \Lambda\tilde{Q}+\tilde{Q}\Lambda^{\tau}= \Lambda\tilde{D}-
  \tilde{D}\Lambda^{\tau}~,
\end{equation}
where
\begin{equation}\label{eq:tilde}
  \tilde{Q} = LQL^{\tau}~, ~~~ \tilde{D}=LDL^{\tau}~.
\end{equation}
 The matrix $\tilde{Q}$ remains antisymmetric and $\tilde{D}$ remains
symmetric. In this representation Eq.\ (\ref{rel4}) takes on the
form
\begin{eqnarray}
  (\lambda_\alpha + \lambda_\beta)\tilde
   Q_{\alpha\beta}+\mu_\alpha\tilde Q_{\alpha + 1,\beta}
   + \mu_\beta\tilde Q_{\alpha,\beta+1}  \nonumber\\
  {\  } {\ } {\ }  =  (\lambda_\alpha-\lambda_\beta)  \tilde
  D_{\alpha\beta}+\mu_\alpha\tilde
  D_{\alpha + 1, \beta} -\mu_\beta\tilde D_{\alpha,\beta + 1}~.
\label{eq:recursion}
\end{eqnarray}
It has the solution, for $\alpha<\beta$,
\begin{eqnarray}
 \tilde Q_{\alpha\beta}
 & = & {\lambda_\alpha-\lambda_\beta\over
\lambda_\alpha+\lambda_\beta}
  \tilde D_{\alpha\beta}   \nonumber\\
 &   & + 2\lambda_\beta\sum_{\mu\ge 1}
{(-1)^{\mu-1}\over(\lambda_\alpha+\lambda_\beta)^{\mu+1}} \tilde
D_{\alpha+\mu,\beta} \nonumber \\
 &   & - 2\lambda_\alpha\sum_{\nu\ge 1}
  {(-1)^{\nu-1}\over(\lambda_\alpha+\lambda_\beta)^{\nu+1}}
  \tilde D_{\alpha,\beta+\nu} \nonumber\\
 &   & \label{eq:fgeneral}
+ 2\sum_{\mu,\nu\ge 1} {(\mu+\nu-1)!\over
\mu!\nu!}{(-1)^{\mu+\nu+1}\over
  (\lambda_\alpha+\lambda_\beta)^{\mu+\nu+1}} \nonumber \\
 &   & (\mu \lambda_\beta-\nu\lambda_\alpha)\tilde D_{\alpha+\mu,\beta+\nu} ~,
\label{rel3}
\end{eqnarray}
 where the sums go over all values of $\mu,\nu$ for which the indices
lie within the same block of the block-diagonal matrix $\Lambda$.
For the case $\lambda_\alpha=\lambda_\beta$ this reduces to
\begin{eqnarray}
  \tilde Q_{\alpha\beta}& = & -\sum_{\mu\ge 1} {(-1)^\mu\over
(2\lambda_\alpha)^\mu}\tilde
  D_{\alpha+\mu,\beta}   + \sum_{\nu\ge 1} {(-1)^\nu\over
  (2\lambda_\alpha)^\nu}\tilde D_{\alpha,\beta+\nu} \nonumber\\
  & & -\sum_{\mu,\nu\ge 1} {(\mu+\nu-1)!\over
\mu!\nu!}{(-1)^{\mu+\nu}\over
  (2\lambda_\alpha)^{\mu+\nu} }  (\mu-\nu) \tilde
D_{\alpha+\mu,\beta+\nu}~. \label{eq:jordandiag}
\end{eqnarray}
For the case in which all the $\mu_\alpha$ are zero, Eq.\
(\ref{eq:fgeneral}) is equivalent to Eq. (\ref{eq:fdiag}).

 In this Jordan representation of force matrix $F$, the state variable
is transformed by
\begin{equation}
  y=Lx~,
\end{equation}
and the corresponding transformation of $U$ is given by
\begin{equation}\label{eq:utilde}
  \tilde U= R^{\tau} U R~.
\end{equation}
Equation (\ref{rel1}) then becomes
\begin{equation}
    (\tilde D+\tilde Q)\tilde U= -\Lambda~,\label{tilde3}
\end{equation}
 which is form-invariant with Eq.~(\ref{rel1}) under the Jordan
transformation.

\newpage 

\end{document}